\documentclass[aps,prl,twocolumn,groupedaddress,nofootinbib]{revtex4}
\usepackage{graphicx}
\usepackage{amsmath}
\usepackage{lineno}
\usepackage{threeparttable}

\begin{document}

\title{Thermal conductivity of MgO, MgSiO$_3$ perovskite and post-perovskite in the Earth's deep mantle}

\author{Volker Haigis$^1$}
\author{Mathieu Salanne$^2$}
\author{Sandro Jahn$^1$}

\affiliation{$^1$ GFZ German Research Centre for Geosciences, Telegrafenberg, 14473 Potsdam, Germany}
\affiliation{$^2$ UPMC Universit\'{e} Paris 06 and CNRS, UMR 7195, PECSA, 75005 Paris, France}

\begin{abstract}
We report lattice thermal conductivities of MgO and MgSiO$_3$ in the perovskite and post-perovskite structures at conditions of the Earth's lower mantle, obtained from equilibrium molecular dynamics simulations. Using an advanced ionic interaction potential, the full conductivity tensor was calculated by means of the Green-Kubo method, and the conductivity of MgSiO$_3$ post-perovskite was found to be significantly anisotropic. The thermal conductivities of all three phases were parameterized as a function of density and temperature. Assuming a Fe-free lower-mantle composition with mole fractions $x_{\mathrm{MgSiO}_3} = 0.66$ and $x_{\mathrm{MgO}} = 0.34$, the conductivity of the two-phase aggregate was calculated along a model geotherm. It was found to vary considerably with depth, rising from 9.5 W/(mK) at the top of the lower mantle to 20.5 W/(mK) at the top of the thermal boundary layer above the core-mantle boundary. Extrapolation of experimental data suggests that at deep-mantle conditions, the presence of a realistic amount of iron impurities lowers the thermal conductivity of the aggregate by about 50\%~\citep{man2011}. From this result and our thermal conductivity model, we estimate the heat flux across the core-mantle boundary to be 10.8 TW for a Fe-bearing MgO/MgSiO$_3$ perovskite aggregate and 10.6 TW for a Fe-bearing MgO/MgSiO$_3$ post-perovskite aggregate.
\end{abstract}

\maketitle

\section{Introduction}
The thermal conductivity of minerals in the Earth's mantle is an important geophysical parameter which governs the heat flux from the core up to the surface and hence strongly influences mantle dynamics~\citep{nal2007}. Moreover, the thermal conductivity of minerals at the core-mantle boundary (CMB) determines the amount of heat extracted from the core, driving the convection of the liquid outer core and thus controlling the power available to the generation of the Earth's magnetic field~\citep{dav2007,aub2009}. Yet, measuring thermal conductivities at mantle pressures and temperatures is extremely challenging, and experimental data are scarce. Several schemes exist to extrapolate thermal conductivities measured at lower pressures and temperatures to deep-mantle conditions~\citep{ros1984,hof1999}, but they are plagued with large uncertainties. Hence a computational approach is desirable to evaluate thermal conductivities directly at the relevant conditions. The aim of this study is to provide reliable values for the lattice thermal conductivities of MgO, MgSiO$_3$ perovskite (Pv) and post-perovskite (PPv) at lower-mantle conditions and their variation with temperature and density (or pressure). These results can be directly applied to thermal transport in the lower mantle.

In deep-mantle minerals, heat is conducted by phonons and electromagnetic radiation. The importance of the radiative contribution to thermal transport in the Earth is under debate, and current estimates span a considerable range: while \cite{gon2009} report a radiative thermal conductivity below $\sim 0.5$ W/(mK) across the lower mantle, \cite{sta2011} predict $\sim 5$ W/(mK) at the CMB, and \cite{kep2008} even values of up to $\sim 10$ W/(mK), which is of the same order of magnitude as the lattice contribution. Moreover, the radiative conductivity seems to depend strongly on crystal grain size and on the iron content \citep{hof2007}. In view of these difficulties, we focus on the lattice contribution in this study. If the radiative conductivity turns out to be significant it can simply be added to the lattice part presented here.

Over the past years, different atomic-scale methods were developed to calculate lattice thermal conductivities. \cite{sta2010} applied the non-equilibrium or ``direct'' method~\citep{mul1997,nie2003} to derive the thermal conductivity of MgO, using molecular dynamics (MD) simulations based on density functional theory (DFT). In this approach, an energy current from the cold to the hot side of the simulation cell is imposed. From this current and the steady-state temperature gradient which builds up, the thermal conductivity is obtained via Fourier's law. While computationally rather efficient, the method suffers from strong finite-size effects, thus requiring extrapolation to infinite system size and introducing considerable uncertainties \citep{sel2010}. An approach based on phonon lifetimes, obtained from DFT, was used by~\cite{dek2009,dek2010} and by~\cite{tan2010} to calculate the thermal conductivity of MgO. Phonon lifetimes were either calculated from line widths in the Fourier transform of the velocity autocorrelation function~\citep{dek2009} or from anharmonic lattice dynamics~\citep{tan2010}. Combined with the Boltzmann transport equation for the phonon gas, they yield the thermal conductivity in the relaxation time approximation. This approach treats the anharmonicity of lattice vibrations perturbatively and is thus limited to temperatures where atomic displacements from the equilibrium positions are small enough for higher-order anharmonicity to be neglected.

A third approach, the Green-Kubo method, uses the Green-Kubo relations~\citep{kub1957} to obtain thermal conductivities from appropriate current correlation functions, which, in turn, are readily extracted from equilibrium MD trajectories. This method has been successfully applied to solids (e.g.~\cite{vol2000, sel2010, esf2011}) and liquids (e.g. \cite{gal2007, oht2009, sal2011}). In contrast to the non-equilibrium method, no concerns about leaving the linear-response regime arise for equilibrium MD. Moreover, the Green-Kubo method exhibits a weaker finite-size effect \citep{sel2010}, provides the full thermal conductivity tensor in one simulation and takes into account thermoelectric effects which can contaminate results of the non-equilibrium method for ionic conductors \citep{sal2011}. Unlike the lattice dynamics approach, the Green-Kubo method takes into account anharmonicity to all orders. Thus its validity is not restricted to low temperatures. In the light of these advantages, we decided to use the Green-Kubo approach to calculate thermal conductivities of MgO, MgSiO$_3$ Pv and MgSiO$_3$ PPv at conditions spanning a wide pressure and temperature range. We also determined conductivities at conditions where experimental data are available, and satisfactory agreement with these experiments makes us confident that our results are equally reliable at CMB conditions. A drawback of the method is that it requires long run durations (in the nanosecond range) to obtain reasonable statistical accuracy. Our calculations are based on classical MD simulations involving an interaction potential of first-principles accuracy \citep{jah2007}.

\section{Theory}
The thermal conductivity tensor $\lambda$ is defined by Fourier's law, $ \mathbf{j}_Q = - \lambda \nabla T$, under the constraint that no mass or electric currents are present. This constraint is relevant to electronic or ionic conductors, where thermoelectric effects occur \citep{cal1985}. Fourier's law is of linear-response type and relates the heat current density $\mathbf{j}_Q$ to the temperature gradient $\nabla T$. For cubic and  orthorhombic crystals, $\lambda$ is diagonal if the coordinate axes are along the crystal axes, and direction-dependent conductivities can be defined by
\begin{equation} \label{eqts1}
 \lambda_{\alpha} = - j_Q^{\alpha} / \nabla_{\alpha} T, \,\,\, \alpha \in \{x,y,z\}
\end{equation}

In the framework of non-equilibrium thermodynamics \citep{cal1985, deg1984}, the thermal conductivity can be expressed in terms of kinetic coefficients $L_{AB}$, as is done in equations \ref{eqts2} and \ref{eqts3} below. They determine the linear response of the system to deviations from equilibrium, i.e. energy and mass flows resulting from thermal and chemical gradients. The gist of the Green-Kubo method is that the kinetic coefficients $L_{AB}$, although representing non-equilibrium behavior, are linked to fluctuations in thermodynamic \textit{equilibrium} via the fluctuation-dissipation theorem. The kinetic coefficients, and hence the thermal conductivity, can therefore be obtained from equilibrium MD by means of appropriate Green-Kubo formulae, which relate the linear response of a system with volume $V$ to current correlation functions in thermodynamic equilibrium:
\begin{equation} \label{eqts4}
 L_{AB}^{\alpha\beta} = \lim_{\tau \rightarrow \infty} \left[ \frac{1}{k_B V} \int_0^{\tau} dt \, \langle J_A^\alpha(t) \, J_B^\beta(0) \rangle \right]
\end{equation}
where $k_B$ is Boltzmann's constant, and the $J_A^\alpha$ are Cartesian components of the energy current ($A=U$) or of the mass currents ($A=1,\dots,N-1$, where $N$ is the number of chemical species in the system), with respective dimensions of energy or mass times velocity. Angular brackets denote an ensemble average. We assume that the center of mass is at rest, hence there are only $N-1$ independent mass currents for a system with $N$ chemical species. Then, for a system with two species, the thermal conductivity is given by \citep{gal2007},
\begin{equation} \label{eqts2}
 \lambda_{\alpha} = \frac{1}{T^2} \left( L_{UU}^{\alpha\alpha} - \frac{(L_{U1}^{\alpha\alpha})^2}{L_{11}^{\alpha\alpha}} \right), \alpha \in\{x,y,z\}
\end{equation}
and for a system with three species by \citep{sal2011}
\begin{equation} \label{eqts3}
\begin{split}
 \lambda_{\alpha} =& \frac{1}{T^2} \left( L_{UU}^{\alpha\alpha} - \right.\\ 
 & \left. \frac{(L_{U1}^{\alpha\alpha})^2 L_{22}^{\alpha\alpha} + (L_{U2}^{\alpha\alpha})^2  L_{11}^{\alpha\alpha} - 2 L_{U1}^{\alpha\alpha} L_{U2}^{\alpha\alpha}  L_{12}^{\alpha\alpha}  }{L_{11}^{\alpha\alpha} L_{22}^{\alpha\alpha} - ( L_{12}^{\alpha\alpha} )^2 } \right)
\end{split}
\end{equation}
It is worth noting that equations \ref{eqts2} and \ref{eqts3} are written here in terms of mass currents, whereas they were originally derived in terms of ionic currents.

\section{Simulation details}
We performed equilibrium molecular dynamics simulations in the $NVT$ ensemble, with a time step of 1 fs for the integration of the equation of motion and a Nos\'{e}-Hoover thermostat \citep{nos1984,hoo1985} maintaining the system at the desired temperature. The cell dimensions were chosen as the average cell size in a previous $NPT$ run at the desired pressure $P$, maintained by a barostat~\citep{mar1994}. The interactions between atoms were described by an advanced ionic interaction potential which was parameterized non-empirically, using DFT as a reference \citep{jah2007}. This potential has been shown to reliably predict properties of minerals of the system CaO-MgO-Al$_2$O$_3$-SiO$_2$ over a wide temperature and pressure range, with accuracy comparable to DFT. In particular, the ionic interaction potential used in this study was shown to describe MgO and the MgSiO$_3$ phases perovskite and post-perovskite well, predicting lattice constants to within 1\% and elastic constants mostly to within 10\%, compared to DFT results~\citep{jah2007}. The elastic constants determine vibrational modes of the crystal in the limit of long wavelengths~\citep{ash1976}. These modes close to the Brillouin zone center, in turn, are expected to make the largest contribution to the thermal conductivity of the crystal~\citep{tan2010}. Therefore, we expect the interaction potential to produce accurate lattice dynamics and thermal transport properties. For MgO, MgSiO$_3$ Pv, and MgSiO$_3$ PPv, we used cubic or orthorhombic supercells containing 512, 960, and 720 atoms, respectively. For each composition, temperature, and pressure, we generated trajectories of at least 0.5 ns and up to 2.4 ns.

\begin{figure}[htp]
\centering
\includegraphics[width=\columnwidth]{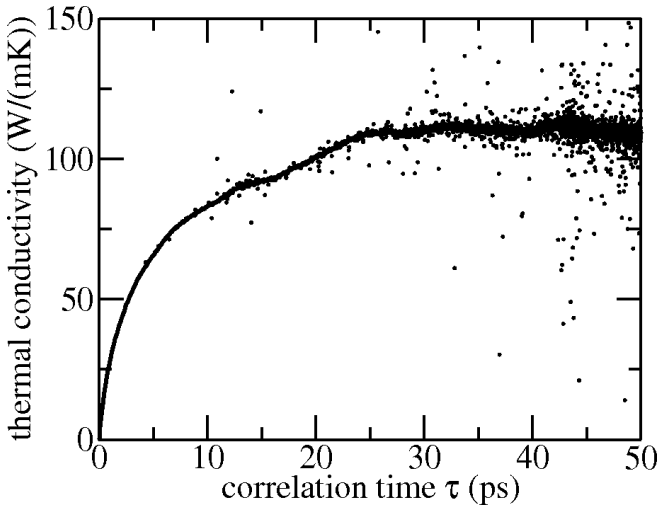}
\caption{Thermal conductivity of MgO at 300 K, 0 GPa as a function of correlation time $\tau$, see eq.~\ref{eqts4}, averaged over 15 MD blocks of 100 ps each. Clearly, a plateau is reached at 30 ps.}
\label{fig1}
\end{figure}

At each time step of the MD run, the mass currents for each species and the energy current were extracted for later calculation of the different current correlation functions needed in eq.~\ref{eqts4}. An explicit expression for the energy current for polarizable ions was derived by~\cite{oht2009}. The total MD run was then divided into blocks of equal length (50 to 100 ps) which were analyzed independently for current correlation functions and thermal conductivity. The correlation time $\tau$ in eq.~\ref{eqts4} was chosen large enough for the thermal conductivity $\lambda$ to reach convergence. In practice, $\lambda$ as a function of $\tau$ oscillates around its limiting value, and for each MD block, we took a time average over the first plateau of the cumulative $\lambda(\tau)$ (averaged over the $\lambda(\tau)$ from the individual MD blocks), see Fig.~\ref{fig1}. Finally, the thermal conductivity and its $1\sigma$ uncertainty were obtained by averaging the results from all blocks.

\section{Results and discussion}
\subsection{MgO}
For reference, we first calculated the thermal conductivity of isotopically pure MgO in the fcc structure at 300 K and ambient pressure. At these conditions, the model predicts a density $\rho$ = 3.602 g/cm$^3$, which is in excellent agreement with the experimental density at ambient conditions, 3.583 g/cm$^3$~\citep{spe2001}. In Fig.~\ref{fig1}, we show the computed thermal conductivity as a function of correlation time $\tau$, see eq.~\ref{eqts4}. The data contain a number of outliers, due to the second term on the right side of eq.~\ref{eqts2}. Both $L_{U1}$ and $L_{11}$ are expected to be close to zero when no diffusion is present, thus the quotient may take on very large positive or negative values occasionally, although it should be small in a crystal. Since outliers tend to distort arithmetic averages, we calculated the conductivity for each MD block from the median of the data over the plateau, which is a more representative measure for the expectation value in such cases.

\begin{table*}
\begin{center}
\caption{Thermal conductivity of fcc MgO, MgSiO$_3$ Pv, and MgSiO$_3$ PPv. $P_{\text{MD}}$ is the pressure resulting from our MD simulations with the indicated density and temperature, and $P_{\text{EoS}}$ the one obtained from an equation of state by~\cite{sti2005}. Results are from this study if not otherwise indicated. For the last $\rho, T$ point of PPv, the three directional conductivities average to (15.3 $\pm$ 0.8) W/(mK), to be compared with the calculated bulk $\lambda$ of (15.1 $\pm$ 0.9) W/(mK). The difference, which is safely within the error bars, results from slightly different averaging schemes: for the bulk value, we first averaged over all directions and then over time (i.e. over the plateau, see fig.~\ref{fig1}), whereas for the direction-dependent conductivities, we averaged each direction individually over the plateau.}
\label{tr1}
\begin{threeparttable}
\begin{tabular}{c c c c c c}
 \hline\hline
 phase & $\rho$ (g/cm$^3$) & $T$ (K) & $P_{\text{MD}}$ (GPa) & $P_{\text{EoS}}$ (GPa) & $\lambda$ (W/(mK)) \\
 \hline
 MgO & 3.602 & 300 & 0 & 1 & 111 $\pm$ 16 \\
 &	&     &   &   & 76 $\pm$ 11 \tnote{a} \\
 &	&     &	  &   & 62 \tnote{b} \\
 &	&     &   &   & 59 $\pm$ 6 \tnote{c} \\
 &	&     &   &   & 75 \tnote{d} \\
 &	&     &   &   & $65 \pm 15$ \tnote{e} \\
 & 5.410 & 300 & 134 & 151 & 1400 $\pm$ 250 \\
 & 4.201 & 2000 & 41 & 45  & 40.0 $\pm$ 2.5 \\
 & 5.307 & 2000 & 133 & 148 & 141 $\pm$ 11 \\
 & 5.307 & 3000 & 138 & 155 & 76.8 $\pm$ 4.4 \\
 \hline
 MgSiO$_3$ Pv & 4.544 & 300 & 26 & 31 & 27.0 $\pm$ 2.2 \\
 &        &     &    & 26  & 19 \tnote{f} \\
 &        &     &    & 31  & 10.6 $\pm$ 0.6 \tnote{g} \\
 & 5.332 & 300 & 107 & 111 & 61.3 $\pm$ 7.9  \\
 &        &     &    & 108.4 & 23.7 $\pm$ 4 \tnote{g} \\
 & 4.544 & 2000 & 40 & 42  & 9.7 $\pm$ 1.0  \\
 & 5.401 & 3000 & 139 & 137 & 12.4 $\pm$ 2.0  \\
 &        & 300 & -- & ambient & 5.1 \tnote{h} \\
 &        & 300 & -- & ambient & 5.8 \tnote{i} \\
 \hline
 MgSiO$_3$ PPv & 5.631 & 298 & 135 & 138 & 167 $\pm$ 25  \\
 &        &      &     & 141 & 65 $\pm$ 14 \tnote{g} \\
 & 5.482 & 2000 & 130 & 132 & 16.8 $\pm$ 0.5 \\
 & 5.631 & 2000 & 150 & 150 & 20.6 $\pm$ 1.7 \\
 & 5.482 & 3000 & 138 & 140 & 15.1 $\pm$ 0.9 \\
 &	&      &       & &$\lambda_x$: 18.0 $\pm$ 1.8 \\
 &	&      &       & &$\lambda_y$: 13.7 $\pm$ 1.0 \\
 &	&      &       & &$\lambda_z$: 14.1 $\pm$ 1.2 \\
 \hline\hline
\end{tabular}
\begin{tablenotes}
\item[a] This study, with isotope correction from~\cite{tan2010}
\item[b] \cite{tan2010}, DFT, with isotope correction
\item[c] \cite{sta2010},  DFT, perfect crystal
\item[d] \cite{dek2010}, DFT, perfect crystal
\item[e] \cite{kat1997}, experiment at ambient conditions
\item[f] \cite{man2011}, experiment at 300 K
\item[g] \cite{oht2012}, experiment at 300 K, pressure determined experimentally
 \item[h] \cite{osa1991}, experiment at 300 K, metastable (quenched to ambient pressure)
\item[i] \cite{oht2012}, high-$P$ experiments, extrapolated to ambient pressure
\end{tablenotes}
\end{threeparttable}
\end{center}
\normalsize
\end{table*}

Our value of the thermal conductivity at ambient conditions is ($111 \pm 16$) W/(mK), significantly larger than that found in other computational and experimental studies (table~\ref{tr1}). However, overestimation with respect to experiments is to be expected, since we considered a perfect, isotopically pure crystal, whereas the experiments were performed on real crystals with natural isotopic composition and defects, which reduces the thermal conductivity considerably relative to its perfect-crystal value \citep{kre2004, tam1983}. Therefore, our results should indeed be larger than the experimental ones. \cite{tan2010} evaluated the isotope effect for MgO and found that at ambient conditions, the thermal conductivity of an isotopically pure crystal exceeds the one of natural samples by as much as 46\%. This correction for isotopic composition is already included in their results in table~\ref{tr1}. If we apply the same correction to our data, we get $\lambda_{\mathrm{MgO}}$(300K,0GPa) = ($76 \pm 11$)W/(mK). Defects, impurities and grain boundaries in real crystals will further reduce the thermal conductivity, and thus our result is fully compatible with the measured conductivity of ($65 \pm 15$) W/(mK)~\citep{kat1997}. On the other hand, the computed values given by \cite{dek2010} and in particular by \cite{sta2010} seem to fall at the low end of values reconcilable with experiments, as the computational data represent isotopically pure, perfect crystals and therefore should \textit{not} agree with conductivities measured on real samples. Following \cite{sel2010}, the relatively small value of \cite{sta2010} may be attributed to the use of a linear extrapolation to account for finite-size effects in the non-equilibrium MD method, which leads to a systematic underestimation of the thermal conductivity.

\begin{figure}[htp]
\centering
\includegraphics[width=\columnwidth]{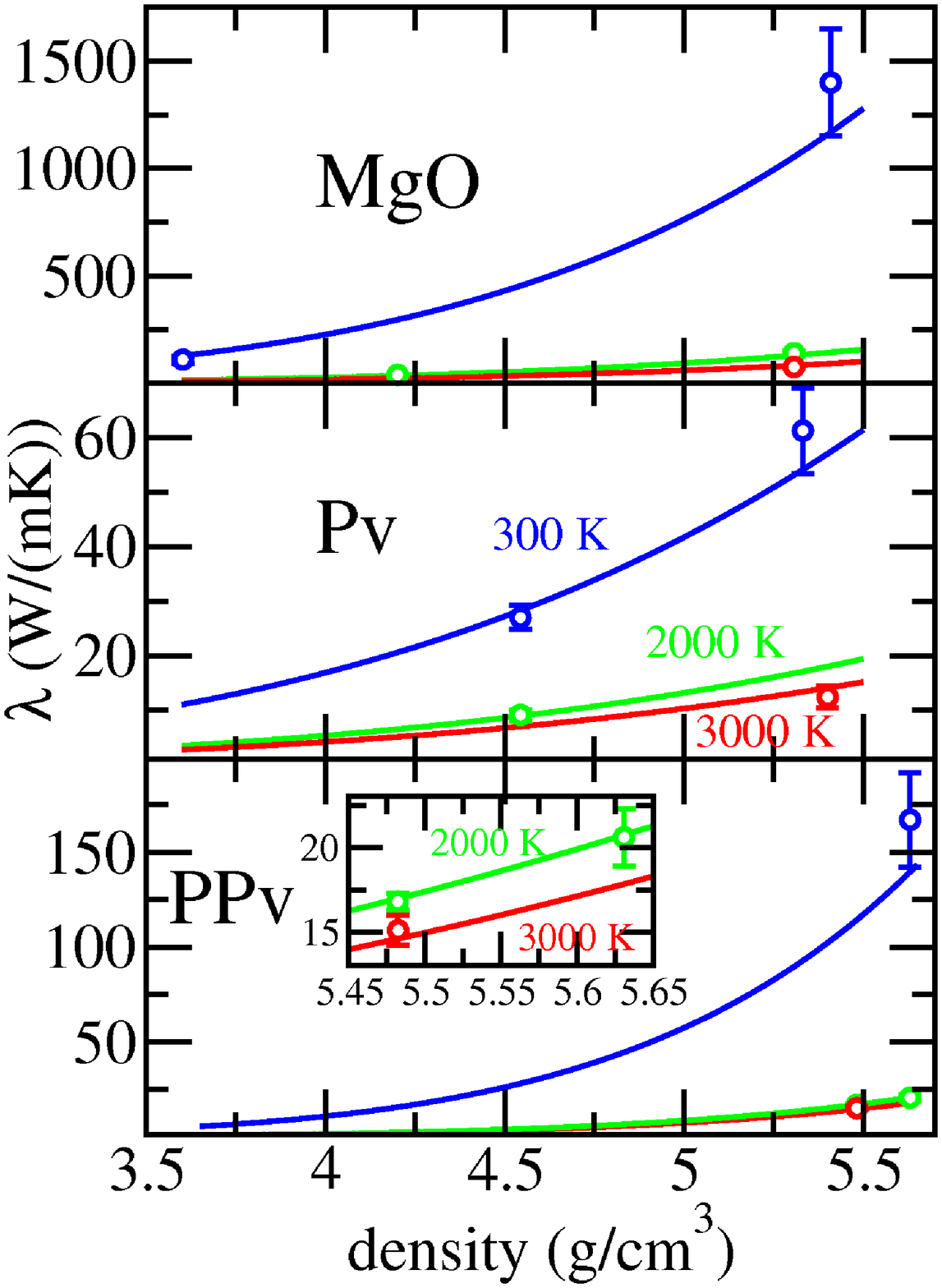}
\caption{Density and temperature dependence of the thermal conductivity for MgO, Pv and PPv. Data points are results of our MD simulations. Lines are fits to eq.~\ref{eqr1} (MgO, Pv) or eq.~\ref{eqr2} (PPv), with colors representing different temperatures. Blue: 300 K, green: 2000 K, red: 3000 K. Inset: PPv at conditions relevant to the lower mantle.}
\label{fig2}
\end{figure}

The thermal conductivity was evaluated at four more $\rho,T$ points, up to lowermost-mantle conditions, $T=3000$ K, $\rho = 5.307$ g/cm$^3$, see table~\ref{tr1}. These data allow us to parameterize the behavior of thermal conductivity over a wide density and temperature range, including the conditions relevant to the lower mantle. The temperature and density dependence of the thermal conductivity is a highly complex matter, and no general theory is currently available \citep{oht2009a}. The dependence on density can be described, in the framework of the Debye approximation, as $\lambda \propto \rho^a$, where $a$ is itself a function of density and temperature in principle \citep{dek2010}. In a recent study, \cite{man2011a} tested the validity of different models for $a(\rho,T)$ by measuring the thermal conductivity of CaGeO$_3$ perovskite, which is an analog phase for MgSiO$_3$. They found subtle differences in the density dependence of $a$ compared to the case of MgO, which they attributed to the larger number of optical phonons in the perovskite phase. However, in view of the limited number of data points in our present study, we take $a$ to be constant, as in \cite{sta2010}, which yields an effective $a$ for the entire density range spanned by our data points. Concerning the temperature dependence, thermal conductivity approximately follows a power law $\lambda \propto T^{-b}$ at high temperatures~\citep{ash1976}. Following \cite{dek2010} and \cite{sta2010}, we write
\begin{equation} \label{eqr1}
 \lambda(\rho,T) = \lambda_{0}\left( \frac{\rho}{\rho_{0}} \right)^a \left( \frac{T_{0}}{T} \right)^b
\end{equation}
with a fixed reference point $\rho_{0} = 3.602$ g/cm$^3$, $T_{0} = 300$ K, and free parameters $\lambda_{0}$, $a$, and $b$. The result of a least-square fit of eq.~\ref{eqr1} to the data points is given in table~\ref{tr2}. The quality of the fit can be assessed by means of fig.~\ref{fig2}.

\begin{table}
\begin{center}
\caption{Parameters for the density and temperature dependence of thermal conductivity resulting from fits to our calculated data points, for MgO in fcc structure, MgSiO$_3$ Pv (both described by eq.~\ref{eqr1}) and PPv, described by eq.~\ref{eqr2}. Reference density $\rho_0$ and temperature $T_0$ are fixed during the fit.}
\label{tr2}
\begin{tabular}{c c c c c c}
 \hline\hline
  & $\rho_0$ (g/cm$^3$) & $T_0$ (K)& $\lambda_0$ (W/(mK)) & $a$ & $b$\\
 \hline
 MgO & 3.602 & 300 & 129 & 5.42 & 1.10 \\
 MgSiO$_3$ Pv & 4.544 & 300 & 28.3 & 4.06 & 0.608 \\
 MgSiO$_3$ PPv & 5.482 & 3000 & 14.6 & 7.48 & 0.327\\
 \hline\hline
\end{tabular}
\end{center}
\end{table}

At the high temperatures prevailing at the CMB, anharmonicity, i.e. phonon-phonon scattering, is expected to be the dominant mechanism limiting the thermal conductivity, compared to other sources of phonon scattering like isotopic disorder, point defects, and grain boundaries. This is because the number of phonons present in the material increases with temperature and hence the mean free path between phonon-phonon collisions becomes shorter than the mean free path imposed by other scattering mechanisms. The reduction of the thermal conductivity due to isotopic disorder in MgO has been shown to decrease from 46\% at room temperature to only 4\% at 4000 K \citep{tan2010}. Qualitatively the same behavior has been observed in experiments on other materials, see e.g. \cite{kre2004}. In MD simulations, anharmonicity, i.e. the dominant scattering mechanism at high temperatures, is automatically included to all orders, and hence our conductivity results for a perfect crystal should be a good approximation for real MgO at CMB conditions. The same argument applies to the MgSiO$_3$ phases to which we turn in the following paragraphs. Although the above reasoning is based on sound physical considerations, we emphasize that more experimental or simulation data are needed to fully understand and quantify the influence of different kinds of defects on the thermal conductivity at high pressure and temperature.

\subsection{MgSiO$_3$ perovskite}
MgSiO$_3$ in the orthorhombic perovskite structure $Pbnm$ is generally accepted to be the most abundant mineral in the Earth's lower mantle (i.e. below a depth of 670 km). It consists of a three-dimensional network of corner-sharing SiO$_6$ octahedra, with Mg occupying the larger inter-octahedral sites. The calculated thermal conductivity of MgSiO$_3$ in the perovskite structure, averaged over all directions, at four state points spanning a wide range of densities and temperatures, is given in table~\ref{tr1}, along with available experimental data. The effect of isotopic disorder on the thermal conductivity is not known for the MgSiO$_3$ phases Pv and PPv. As in the case of MgO, it may be significant at low temperatures but is expected to decrease rapidly with temperature. The density and temperature dependence of thermal conductivity is well described by eq.~\ref{eqr1}, and the respective fit parameters are listed in table~\ref{tr2}. In fig.~\ref{fig2}, the model conductivity is plotted along with the computed data points.

The available experimental data for perovskite scatter considerably (table~\ref{tr1}), and the effect of grain boundaries, isotopic disorder and other defects like possible cracks in the samples is not known quantitatively. Hence a comparison to our results is difficult. As expected, the results obtained from perfect-crystal simulations are larger than the experimental values. Our results seem compatible with those of \cite{man2011} but more difficult to reconcile with the data of \cite{oht2012}. Note that at 300 K and approximately 30 GPa, \cite{oht2012} report a considerably lower conductivity than \cite{man2011}: the data, both derived from measurements, differ by a factor of almost 2. At ambient conditions, \cite{osa1991} report the conductivity to be 5.1 W/(mK), which is in line with the value of 5.8 W/(mK), derived from a fit to experimental data by \cite{oht2012}.

\subsection{MgSiO$_3$ post-perovskite}
The perovskite structure of MgSiO$_3$ transforms to an orthorhombic post-perovskite phase ($Cmcm$) at approximately 125 GPa and 2500 K~\citep{mur2004,oga2004} which is believed to be stable in the Earth's lowermost mantle close to the core-mantle boundary and might be responsible for the D$^{\prime\prime}$ seismic discontinuity \citep{iit2004}. It is characterized by layers of corner- and edge-sharing SiO$_6$ octahedra perpendicular to the $b$ axis, with Mg occupying inter-layer sites. This anisotropic structure exhibits strongly anisotropic elastic properties~\citep{iit2004}, which should lead to direction-dependent phonon velocities, and hence we expect anisotropic thermal transport properties. Therefore, in addition to the direction-averaged conductivity, we also calculated the thermal conductivities separately along the three axes of the orthorhombic crystal for one data point, at conditions representative of the lowermost mantle. The simulation time was extended to 2.35 ns in this case to ensure satisfying statistics for each direction individually. The results for several $\rho,T$ values are listed in table~\ref{tr1}.

The calculated data points could not be fit satisfactorily with eq.~\ref{eqr1}, due to the very weak temperature dependence of the thermal conductivity above 2000 K (compare the second and fourth data point of PPv in table~\ref{tr1}). This flattening of the thermal conductivity as a function of temperature is consistent with experimental observations which show a near-constant conductivity above a certain temperature, depending on the material studied \citep{per2008}. In parameterizing the thermal conductivity of PPv, we therefore used an expression which reconciles a strong temperature dependence at lower temperatures with a flat behavior at high temperatures. A good fit could be obtained with the following functional form:
\begin{equation} \label{eqr2}
 \lambda(\rho,T) = \lambda_{0}\left( \frac{\rho}{\rho_{0}} \right)^a \left( \frac{T_{0}}{T - 295 \text{K}} \right)^b \left( \frac{T_0 - 295 \text{K}}{T_0} \right)^b
\end{equation}
where $\rho_{0}$ and $T_{0}$ are fixed reference values and $\lambda_{0}$, $a$, and $b$ are fitting parameters, and the last (normalizing) factor ensures that $\lambda(\rho_0,T_0) = \lambda_0$. The results of a least-square fit are listed in table~\ref{tr2} and compared to the data in fig.~\ref{fig2}. Due to the somewhat empirical nature of the assumed temperature dependence, the validity of eq.~\ref{eqr2} is restricted to the temperature range covered by our data points. It can certainly not be applied below 298 K: in fact, the expression diverges at $T=295$ K. However, we stress that all data points are well fitted. In particular, the density and temperature dependence at conditions of the lower mantle is well captured by the model, as can be seen from the inset in fig.~\ref{fig2}.

The calculated conductivity at conditions of the lowermost mantle is clearly anisotropic, and it is lowest in the $y$ direction (along the $b$ axis of the crystal). This is consistent with the fact that the crystal is softer along $b$ (perpendicular to the layers formed by corner- and edge-sharing SiO$_6$ octahedra) than along $a$ and $c$ (in the plane of the SiO$_6$ sheets). This leads to lower phonon velocities along $b$, at least close to the Brillouin zone center, and a reduced conductivity.

The thermal conductivity of (167 $\pm$ 25) W/(mK), obtained at 300 K and $\rho = 5.631$ g/cm$^3$, is considerably higher than the value derived from experiments at similar conditions by \cite{oht2012}, which is ($65 \pm 14$) W/(mK) (table~\ref{tr1}). Although we cannot quantify the effect of defect scattering in their polycrystalline PPv sample of natural isotopic composition, the discrepancy seems too large to be completely explained by this mechanism, and its origin remains unclear. We note however, that in the case of perovskite, the thermal conductivity reported by \cite{oht2012} was much lower than the experimental value by \cite{man2011}. Interestingly, the estimate of \cite{oht2012} for the thermal conductivity of PPv at 3000 K and 135 GPa, based on an assumed temperature dependence, is close to and even slightly higher than our result at similar conditions (19.5 W/(mK) and 15.1 W/(mK), respectively). This agreement at high $T$ may partially be due to a fortuitous cancellation of discrepancies, since \cite{oht2012} assumed a different temperature dependence than the one we found. But it also hints at the fact that differences between perfect-crystal simulations and real-sample experiments become less important with increasing temperature.

\section{Implications for the thermal conductivity of the Earth's lower mantle}
When applying our results to heat transport in the Earth's deep mantle, they should be considered upper estimates for the lattice thermal conductivities of real minerals. Our calculations do not take into account the natural isotopic composition, impurities, and defects of the minerals, all of which lower the conductivity. \cite{man2011} measured the effect of realistic amounts of iron impurities on the thermal conductivity of MgO and MgSiO$_3$ perovskite, at relatively low temperatures and pressures. By extrapolation, they estimate that 20 mol\% and 3 mol\% of iron in MgO and perovskite, respectively, reduce the thermal conductivity of the aggregate at the CMB by about 50\% relative to that of the chemically pure aggregate. On the other hand, we did not take into account radiative heat transport as an additional mechanism of thermal conduction.

To calculate the thermal conductivity in the lower mantle, we assumed a simplified mantle composition, derived from the pyrolitic composition given by \cite{pia2007}, with molar fractions $x_{\mathrm{MgSiO}_3} = 0.66$ (perovskite structure) and $x_{\mathrm{MgO}} = 0.34$. The conductivities were calculated along a mantle geotherm, with the depth-dependent pressure taken from the Preliminary Reference Earth Model \citep{dzi1981} and the temperature profile adopted from \cite{sta2008}. Since our model for thermal conductivities was parameterized as a function of density and temperature, the pressures along the geotherm had to be converted to densities. This was done by means of equations of state for the mineral phases, described by \cite{sti2005}, with a revised set of parameters for that model taken from \cite{xu2008}. The pressures resulting from the equations of states for a given density and temperature agree well with the pressures obtained directly from our MD simulations (table~\ref{tr1}). This further corroborates the adequacy of the interaction potential used in the simulations in describing material properties over the $\rho,T$ range of the lower mantle. By means of the equations of state, the mole fractions of the individual mineral phases can be converted to volume fractions, yielding about 82\% Pv and 18\% MgO by volume in the lower mantle. These numbers change slightly with pressure and temperature, and the exact values resulting from the equations of state have been used throughout the study.

The thermal conductivity of a two-phase aggregate, expressed in terms of the conductivities, $\lambda_1, \lambda_2$, and volume fractions, $f_1, f_2,(f_1+f_2=1)$, of the individual phases, depends on the geometric details of the assemblage. The extreme cases are realized by a structure of alternating parallel layers of the two phases, with a heat flux parallel and perpendicular to the layers, respectively. For the former case (a ``parallel circuit''), the conductivity of the aggregate is maximum and given by the arithmetic or Voigt average $\lambda_{\mathrm{max}} = f_1 \lambda_1 + f_2 \lambda_2$. For the latter (a ``series circuit''), the conductivity takes on its minimum or Reuss average $\lambda_{\mathrm{min}} = \left( f_1/\lambda_1 + f_2/\lambda_2 \right)^{-1}$. For other geometries, not necessarily built from layers, the conductivity of the aggregate will lie within these bounds.

\begin{figure}[htp]
\centering
\includegraphics[width=\columnwidth]{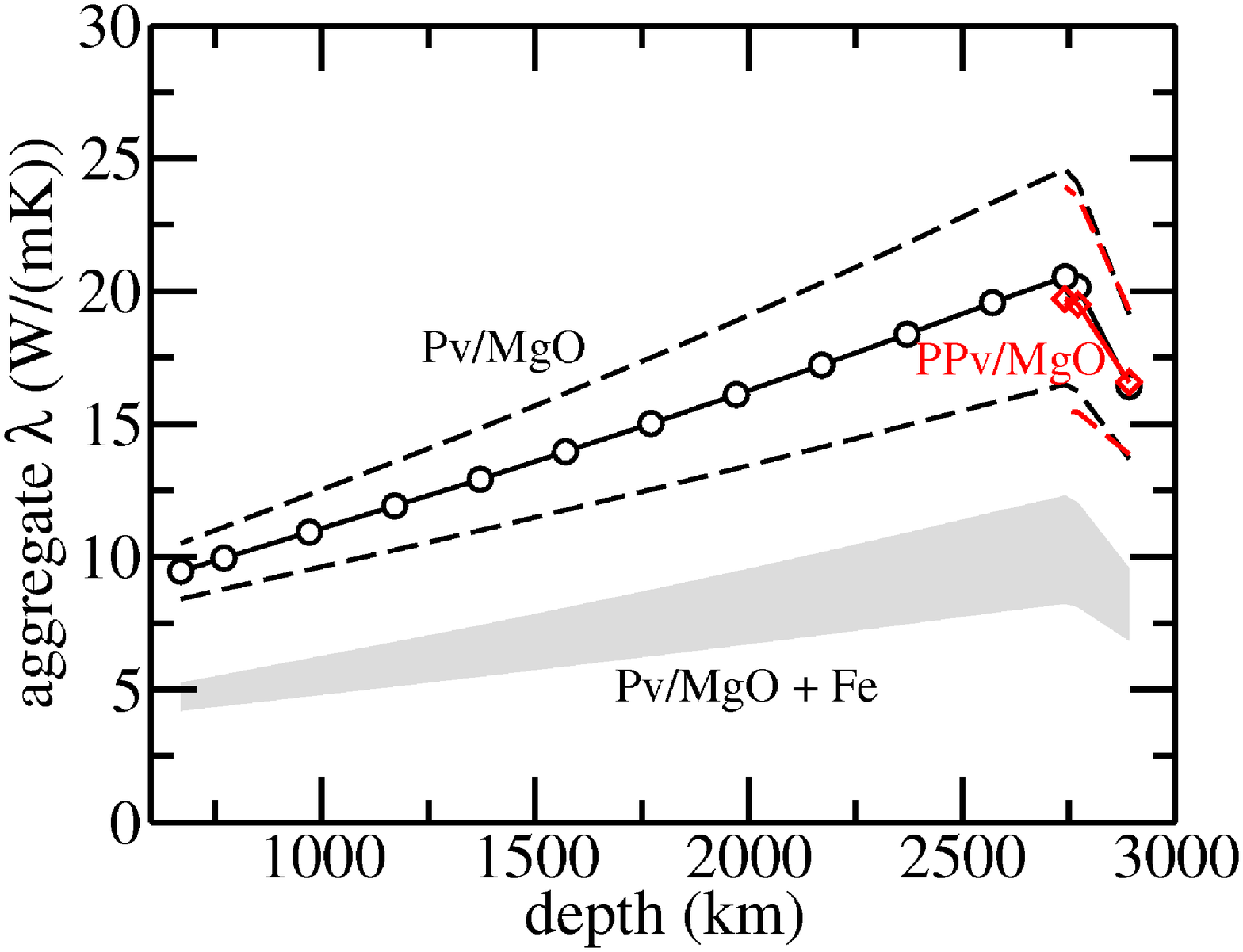}
\caption{Average thermal conductivity $\overline{\lambda}$ along a model geotherm \citep{sta2008} for an aggregate with mole fractions $x_{\mathrm{MgSiO}_3} = 0.66$ (black circles: perovskite structure, red diamonds: post-perovskite strucutre) and $x_{\mathrm{MgO}} = 0.34$. The dashed lines represent $\lambda_{\mathrm{min}}$ and $\lambda_{\mathrm{max}}$, bracketing the geometry-dependent aggregate conductivity (see text). Lines are a guide to the eye. The grey-shaded area indicates the range of the Pv/MgO aggregate thermal conductivity under the assumption that iron in the Earth's lower mantle reduces the conductivity of the aggregate by 50\% \citep{man2011}.}
\label{fig3}
\end{figure}

Fig.~\ref{fig3} shows the thermal conductivity of the MgSiO$_3$(Pv)-MgO aggregate along the model geotherm and, for comparison, of the MgSiO$_3$(PPv)-MgO aggregate close to the CMB. Thermal conductivity increases with pressure and decreases with increasing temperature, see eq.~\ref{eqr1}. With increasing depth, the pressure effect dominates over the concomitant temperature rise, resulting in a net increase of the thermal conductivity. Only close to the CMB, the sharp rise in temperature in the thermal boundary layer reverses this trend. At 2891 km depth, i.e. at the CMB, with $P=136$ GPa and $T=3739$ K, the conductivity of a MgO/MgSiO$_3$ perovskite aggregate is predicted to lie between $\lambda_{\mathrm{min}} = 13.7$ W/(mK) and $\lambda_{\mathrm{max}} = 19.1$ W/(mK), depending on geometry, with average $\overline{\lambda} = 16.4$ W/(mK). With MgSiO$_3$ in the post-perovskite structure instead, we obtain $\lambda_{\mathrm{min}} = 13.9$ W/(mK), $\lambda_{\mathrm{max}} = 19.3$ W/(mK), and $\overline{\lambda} = 16.6$ W/(mK), i.e. changes are not significant. The value for the PPv/MgO aggregate is in good agreement with \cite{oht2012}, who estimate the aggregate conductivity at 4000 K and 135 GPa to be approximately 16 W/(mK). Assuming that a realistic amount of iron impurities reduces the aggregate conductivities by 50\% \citep{man2011}, the respective average conductivities at the CMB are 8.2 W/(mK) for the MgO/perovskite and 8.3 W/(mK) for the MgO/post-perovskite aggregate. The influence of iron on the thermal conductivity is treated approximatively here, and more data are needed to better quantify this effect at high pressures and temperatures.

Our results for the thermal conductivity across the lower mantle for an iron-free composition are in remarkable agreement with (albeit slightly larger than predicted by) the thermal conductivity model by \cite{man2011} which is based on an extrapolation of low-$P,T$ experimental data to CMB conditions, assuming an aggregate of 20\% MgO and 80\% Pv by volume. On the other hand, the model used by \cite{hof1999} to extrapolate available conductivity data to high $P,T$ (including iron) yields somewhat smaller values, ranging approximately between 4 W/(mK) and 7 W/(mK) across the lower mantle. Also the estimate of the lower-mantle thermal conductivity by \cite{gon2009} lies below our data: for the lattice conductivity, they assumed an iron-free mantle composition (20\% MgO and 80\% Pv by volume) and extrapolated experimental data, finding a maximum lattice thermal conductivity $\lambda_{\mathrm{max}}$ varying from approximately 3 W/(mK) to 11 W/(mK) across the lower mantle. \cite{sta2011}, using the Ross model \citep{ros1984} and approximative equations of state to derive high-$P,T$ lattice thermal conductivities from available data, also obtained values lower than ours, with $\lambda_{\mathrm{max}}$ not exceeding 8 W/(mK) for an iron-free mantle composition (20\% MgO and 80\% Pv by volume). We emphasize that our data are based on simulations directly at lower-mantle conditions and do not depend on extrapolations.

Finally, an estimate for the heat flux across the CMB is presented. Given the temperatures $T_1$ and $T_2$ at two different depths $z_1$ and $z_2$ as boundary conditions, the steady-state heat current density $j_Q$ is determined by an integral form of Fourier's law,
\begin{equation*}
 j_Q = -\frac{1}{z_2 - z_1} \int_{T_1}^{T_2} dT \lambda(\rho,T)
\end{equation*}
where the two concentric spheres corresponding to $z_1$ and $z_2$ are locally approximated as parallel planes. Taking the temperature at the CMB and 120~km above from \cite{sta2008} and neglecting the small density variation across this layer, we obtain an average CMB heat flux of 21.5 TW for a Pv/MgO aggregate and of 21.2 TW for a PPv/MgO aggregate. This estimate is based on a specific thermal model of the Earth, and a different temperature profile at the CMB would lead to a somewhat different estimate of the CMB heat flux. Assuming that the presence of iron impurities reduces the heat flux by 50\% \citep{man2011}, it is estimated to be 10.8 TW on average for an Fe-bearing Pv/MgO aggregate and 10.6 TW for a Fe-bearing PPv/MgO aggregate, with possible variations by about $\pm$20\%, depending on the geometric details of the two-phase assemblage. These values for the CMB heat flux are consistent with previous estimates, spanning a wide range from 5 TW to 15 TW \citep{lay2008}.

\section{Conclusions}
We performed equilibrium MD simulations and used the Green-Kubo method to calculate lattice thermal conductivities of MgO, MgSiO$_3$ perovskite, and MgSiO$_3$ post-perovskite over a wide range of pressure and temperature conditions relevant to the Earth's deep mantle. To our knowledge, these are the first simulation results for the MgSiO$_3$ phases. Moreover, the thermal conductivity of the lowermost mantle has been determined directly, without extrapolation from experimental or computational low-pressure or low-temperature data and hence is free of the inherent uncertainties.

The data were used to construct a model for thermal conductivities as a function of density and temperature, which was then applied to the Earth's lower mantle. The thermal conductivity was found to increase significantly with depth and to decrease steeply across the thermal boundary layer above the CMB. These results may be used in geodynamic modeling to refine large-scale simulations of mantle convection. In this field, one often assumes a constant thermal diffusivity $\lambda/(\rho c_P)$ ($\rho$: densitity, $c_P$: specific heat capacity) across the mantle (e.g., \cite{tan2011}), which is poorly constrained, moreover. Together with the approximation $c_P = $ const., this implies $\lambda \propto \rho$, a rather restrictive assumption, to be contrasted with the more flexible eq.~\ref{eqr1}.

By combining our thermal conductivity results with a thermal model of the Earth \citep{sta2008}, the lattice contribution to the CMB heat flux is estimated to be about 11 TW for a Fe-bearing two-phase aggregate (virtually the same with MgSiO$_3$ perovskite and post-perovskite). This relatively high flux is consistent with recent estimates of the heat flux required to generate and maintain mantle plumes \citep{lay2008}. Due to the large conductivity contrast between MgO and the MgSiO$_3$ phases, the conductivity of the two-phase aggregate depends strongly on the aggregate geometry. Thus, the CMB heat flux may show large lateral variations by up to about $\pm 20$\%.

\section{Acknowledgments}
We thank Thomas C. Chust for the numerical evaluation of the equations of state as well as Hauke Marquardt, Sergio Speziale and Rene Ga\ss{}m\"oller for fruitful discussions. V.H. and S.J. acknowledge financial support of the Deutsche Forschungsgemeinschaft (DFG) through the Grant No. JA1469/4-1 from the Emmy-Noether-Program. Part of the work was carried out under the HPC-EUROPA2 project (project number: 228398) with the support of the European Commission Capacities Area - Research Infrastructures Initiative. We also acknowledge support of DAAD-PROCOPE under grant no. D/9811428.

\end{document}